\documentclass[aps, prd, twocolumn, nofootinbib]{revtex4}

\usepackage{graphicx}

\begin{document}

\title{Inflation over the hill}

\author{Konstantinos Tzirakis} 
\email{ct38@buffalo.edu}
\author{William H.  Kinney} 
\email{whkinney@buffalo.edu}
\affiliation{Dept. of Physics, University at Buffalo, the  State University of New York, Buffalo, NY 14260-1500}

\begin{abstract}
We calculate  the power spectrum  of curvature perturbations  when the
inflaton  field is  rolling  over the  top  of a  local  maximum of  a
potential.  We show that the  evolution of the field can be decomposed
into  a late-time  attractor, which  is  identified as  the slow  roll
solution,   plus   a   rapidly   decaying  non-slow   roll   solution,
corresponding to the  field rolling ``up the hill''  to the maximum of
the potential.  The exponentially  decaying transient solution can map
to an observationally relevant range of scales because the universe is
also expanding exponentially.  We consider the two branches separately
and we find  that they are related through  a simple transformation of
the  slow  roll parameter  $\eta$  and  they  predict identical  power
spectra.  We generalize  this approach to the case  where the inflaton
field is described  by both branches simultaneously and  find that the
mode equation  can be  solved exactly at  all times.  Even  though the
slow roll  parameter $\eta$ is evolving rapidly  during the transition
from the  transient solution to the late-time  attractor solution, the
resultant  power  spectrum  is  an  exact  power-law  spectrum.   Such
solutions may be useful for model-building on the string landscape.

\end{abstract}

\pacs{98.80.Cq}

\maketitle

\section{Introduction}

The  inflationary  paradigm \cite{Guth:1980zm}  is  one  of the  basic
constituents  of the  standard cosmological  model.  By  postulating a
short period of  accelerated expansion of the very  early universe, it
provides a natural explanation for  the origin of structure as well as
the  flatness and  homogeneity  of the  observable  universe.  In  the
simplest models,  inflation is driven  by a single scalar  field ({\it
inflaton}) which  evolves slowly along  a nearly flat  potential.  The
quantum  fluctuations of  the  inflaton can  then  be translated  into
perturbations in the early  universe which are observed as temperature
fluctuations  in  the  cosmic  microwave background  (CMB)  radiation.
These   models  predict  an   almost  scale-invariant   spectrum  with
negligible non-gaussianities for the primordial density perturbations.

Unfortunately, the equations of motion  for the inflaton field can not
be solved  analytically in general and certain  approximations must be
made  in   order  to  obtain  exact  solutions.    The  most  powerful
approximation  is the  slow roll  approximation \cite{Albrecht:1982wi,
Steinhardt:1984jj,  Linde:1981mu},  which  assumes  that  the  kinetic
energy of  the inflaton field is  suppressed by the  expansion, and at
higher order amounts  to the introduction of an  infinite hierarchy of
parameters  \cite{Liddle:1994dx}.  It  is  then possible  not only  to
obtain analytic solutions of the  equation of motion of the field, but
also  to  express  the  power  spectrum  in terms  of  the  slow  roll
parameters.

Even  though  the  slow  roll  approximation is  widely  used  in  the
literature,  it is  by no  means  the only  successful description  of
inflation. It is well known that the slow roll approximation, while it
initially  describes the evolution  of the  field with  great accuracy,
breaks  down  near the  end  of  inflation.   Moreover, one  can  find
solutions   in   cases   where   slow  roll   is   strongly   violated
\cite{Garcia-Bellido:1996ke,       Stewart:1993bc,       Linde:2001ae,
Kinney:1997ne,   Easther:1995pc,  Kinney:2005vj}.   It   is  therefore
interesting to investigate inflation  in situations where, even though
the  slow roll  approximation  is  not valid,  the  power spectrum  of
curvature perturbations agrees with observations.

The paper  is structured as follows:  Section II presents  a review of
the single field  inflation formalism.  In section III  we discuss the
generation  of curvature  perturbations in  inflation and  we  find an
exact expression  for a  specific class of  models.  In section  IV we
study the case  of the inflaton field rolling over the  top of a local
maximum of a potential.   We show  that even  though the  slow roll
parameter $\eta$  is evolving  rapidly and the  inflationary solutions
are  initially far from  slow roll,  the mode  equation can  be solved
exactly   and  the   generated  perturbations   are   consistent  with
observations. We  apply this  exact solution to  the case  of non-slow
roll  evolution considered  in Ref.  \cite{Kinney:2005vj}.   Section V
contains a summary and conclusions.

\section{The single-field inflation formalism}

In  this  section  we  review  the basic  formalism  of  single  field
inflation. The  background metric that will  be used in  this paper is
the flat FRW metric
\begin{equation}
\label{background metric}
ds^2 = dt^2-a^2(t)d \mathbf{x}^2=a^2(\tau)[d\tau^2-d \mathbf{x}^2],
\end{equation}
where $\tau$ is the conformal time, defined in terms of the coordinate
time $t$ as
\begin{equation}
\label{tau}
d\tau=\frac{1}{a}dt.
\end{equation}
In what  follows, overdots correspond  to derivatives with  respect to
the coordinate time $t$, and primes to derivatives with respect
to  the field  $\phi$. Assuming  that the  field $\phi$  dominates the
energy density of the universe, the Einstein field equations become
\begin{eqnarray}
\label{field equations}
H^2=\left(\frac{\dot a}{a}\right)^2 & = &  \frac{8 \pi}{3 m_{\rm Pl}^2}\left[V(\phi)+\frac{1}{2}\dot  \phi^2\right],\nonumber\\ 
\frac{\ddot a}{a} & = & \frac{8 \pi}{3 m_{\rm Pl}^2}\left[V(\phi)-\dot \phi^2\right].
\end{eqnarray}
The evolution  of the scale  factor $a$ is  then given by  the general
expression
\begin{equation}
\label{scale factor}
a \propto e^{-N},
\end{equation}
where $N$ is defined to be the number of e-folds
\begin{equation}
\label{N}
dN \equiv -H dt.
\end{equation}
According  to the  above  sign  convention, $N$  gives  the number  of
e-folds  before  the  end  of  inflation and  increases  as  one  goes
backwards in time.

The equation of motion for a spatially homogeneous scalar field $\phi$
in a FRW background can be written
\begin{equation}
\label{EoM}
\ddot \phi + 3 H \dot \phi + V^{\prime} (\phi)=0.
\end{equation}
In the  case that $\ddot  \phi$ is negligible, we
recover the slow roll approximation
\begin{equation}
\label{SR EoM}
3 H \dot \phi + V^{\prime} (\phi) \simeq 0,
\end{equation}
with
\begin{equation}
H^2 \simeq {8 \pi \over 3 m_{\rm Pl}^2} V\left(\phi\right).
\end{equation}
If the  field $\phi$ is monotonic  in time, we can  express the Hubble
parameter $H$ in  terms of $\phi$ instead of  the coordinate time $t$.
This  is in  general more  convenient since  $H$ is  not  constant but
varies  as the  field evolves  along the  potential.  The  dynamics of
single  field  inflation  can  then  be  described  by  the  so-called
Hamilton-Jacobi   formalism   \cite{Muslimov:1990be,   Salopek:1990jq,
Lidsey:1995np}, given by the following equation
\begin{equation}
\label{HJ}
\left[H^{\prime}  (\phi)\right]^2-\frac{12  \pi}{m_{\rm Pl}^2}H^2(\phi)  =
-\frac{32\pi^2}{m_{\rm Pl}^4}V(\phi),
\end{equation}
and
\begin{equation}
\label{intro dphi dt 1}
\dot{\phi}=-\frac{m_{\rm Pl}^2}{4 \pi}H'(\phi).
\end{equation}
This system of  two first order equations is  equivalent to the second
order equation of motion (\ref{EoM}).

We  can then  define  the  following Hubble  slow  roll parameters  as
derivatives of $H$ with respect to the field $\phi$
\begin{equation}
\label{intro eps 1}
\epsilon \equiv \frac{m_{\rm Pl}^2}{4\pi}\left(\frac{H^{\prime}(\phi)}{H(\phi)}\right)^2 \simeq
\frac{m_{\rm Pl}^2}{16\pi}\left(\frac{V^{\prime}(\phi)}{V(\phi)}\right)^2,
\end{equation}
\begin{equation}
\label{intro eta 1}
\eta \equiv \frac{m_{\rm Pl}^2}{4\pi}\frac{H^{\prime\prime}(\phi)}{H(\phi)} \simeq \frac{m_{\rm Pl}^2}{8\pi}\left[\frac{V^{\prime\prime}(\phi)}{V(\phi)}-\frac{1}{2}\left(\frac{V^{\prime}(\phi)}{V(\phi)}\right)^2\right],
\end{equation}
and
\begin{equation}
\label{xi}
\xi^2 \equiv \frac{m_{\rm Pl}^4}{16 \pi^2}\frac{H^{\prime}(\phi) H^{\prime
\prime \prime}(\phi)}{H^2(\phi)}.
\end{equation}
The last terms in Eqs. (\ref{intro eps 1}) and (\ref{intro eta 1}) are
true only  in the context of  slow roll, in which  case $\epsilon$ and
$\eta$ are  functions of the  derivatives of the potential.   In order
for slow  roll to be valid, the slope  and the curvature of
the potential should satisfy $V^{\prime}(\phi)
\ll  V(\phi)$   and  $V^{\prime  \prime}(\phi)   \ll  V(\phi)$,
or in terms  of the slow roll parameters $\epsilon \ll 1$ and $|\eta| \ll 1$.

We next derive identities  which will be
useful in  the following discussion.  Using Eq.
(\ref{intro dphi dt 1}) it can be shown that
\begin{equation}
\label{H '' wrt phi}
H''(\phi)=-\frac{4 \pi}{m_{\rm Pl}^2}\frac{\ddot \phi}{\dot \phi}.
\end{equation}
Substituting back into Eq. (\ref{intro eta 1})
we find that
\begin{equation}
\label{intro eta 2}
\eta = 3 + \frac{V^{\prime}(\phi)}{H \dot\phi},
\end{equation}
where Eq. (\ref{EoM})  was used.  It should be  noted that even though
the above  expression contains the first derivative  of the potential,
it is exact.

The slow roll  parameters $\epsilon$ and $\eta$ can  also be expressed
in  terms  of  the  field  $\phi$  and  the  number  of  e-folds  $N$.
Substituting Eq. (\ref{intro  dphi dt 1}) into (\ref{intro  eps 1}) we
find that
\begin{equation}
\label{intro eps 2}
\epsilon=\frac{4 \pi}{m_{\rm Pl}^2}\left(\frac{\dot \phi}{H}\right)^2.
\end{equation}
Also from Eq. (\ref{N}) it is obvious that
\begin{equation}
\label{intro dphi dt 2}
\dot \phi =-H \frac{d\phi}{dN},
\end{equation}
and we then have
\begin{equation}
\label{intro eps 3}
\epsilon=\frac{4 \pi}{m_{\rm Pl}^2}\left(\frac{d\phi}{dN}\right)^2.
\end{equation}
We    can    similarly   express    $\eta$    as    a   function    of
$\phi$. Differentiating Eq. (\ref{intro eps 3})
with respect to $N$
\begin{equation}
\label{intro eps 4}
\frac{d\epsilon}{d\phi}=\frac{8 \pi}{m_{\rm Pl}^2}\frac{d^2\phi}{dN^2},
\end{equation}
where \cite{Kinney:2002qn}
\begin{equation}
\label{deps dphi}
\frac{d\epsilon}{d\phi}=\frac{4\sqrt{\pi}}{m_{\rm Pl}}\sqrt{\epsilon}(\eta-\epsilon).
\end{equation}
Equating  the  above  two   expressions  and  substituting  back  into
 Eq. (\ref{intro eps 3}), we find that
\begin{equation}
\label{intro eta 5}
\eta=\left(\frac{d^2\phi}{dN^2}\right)\left(\frac{d\phi}{dN}\right)^{-1}+\epsilon.
\end{equation}
In  the  next  section,  we  summarize  the  generation  of  curvature
perturbations in  inflation and  we find an  exact expression  for the
curvature power spectrum in the special case where $\epsilon$ is small
and $\epsilon \ll |\eta|$=constant.

\section{Curvature perturbations in inflation}

Inflation provides a natural explanation for the large-scale structure
of  the  universe.   During  inflation, quantum  fluctuations  of  the
inflaton field  are stretched  to scales much  larger than  the horizon
size,             creating             metric            perturbations
\cite{Mukhanov:1981xt,Hawking:1982my,Guth:1982ec,Bardeen:1983qw}.
These   perturbations  are   of   two  kinds:   scalar  or   curvature
perturbations,  which  are responsible  for  structure formation,  and
tensor  perturbations (gravitational  waves).  In  this paper  we will
only consider scalar perturbations, which can be described using
the gauge-invariant variable $u$ \cite{Mukhanov:1990me}\,

\begin{equation}
\label{gauge inv}
u=a \delta \phi-\frac{a \dot \phi}{H}\mathcal{R},
\end{equation}
where  $\mathcal  {R}$  is  the  metric  curvature  perturbation.   On
comoving hypersurfaces  ($\delta \phi=0$), the  curvature perturbation
$\mathcal{R}$ can be written
\begin{equation}
\label{R}
\mathcal{R} = \left|\frac{u}{z} \right|,
\end{equation}
where $z$ is defined as
\begin{equation}
\label{z}
z = \frac{a \dot \phi}{H}.
\end{equation}
We then  define the  power spectrum of  $\mathcal{R}$ in terms  of its
two-point correlation function
\begin{equation}
\label{P(k) R}
P^{1/2}_{\mathcal{R}}(k)=\left[\frac{k^3}{2\pi^2} \langle \mathcal{R}^2 \rangle \right]^{1/2}.
\end{equation}
Using Eq. (\ref{R}),   we   can   express
$P_{\mathcal{R}}(k)$ as
\begin{equation}
\label{P(k) u}
P^{1/2}_{\mathcal{R}}(k)=\sqrt{\frac{k^3}{2
\pi^2}}\left|\frac{u_{k}}{z}\right|,
\end{equation}
where   the    mode   function   $u_{k}$    satisfies   the   equation
\cite{Mukhanov:1985rz, Mukhanov:1988jd}
\begin{equation}
\label{mode eq 1}
\frac{d^2u_{k}}{d\tau^2}+\left(k^2-\frac{1}{z}\frac{d^2z}{d
\tau^2}\right)u_{k}=0,
\end{equation}
and
\begin{equation}
\label{z term}
\frac{1}{z}\frac{d^2z}{d\tau^2}=2a^2H^2\left(1+\epsilon+\epsilon^2-\frac{3}{2}\eta+\frac{1}{2}\eta^2-2\epsilon\eta+\frac{1}{2}\xi^2\right).
\end{equation}
It should be noted that even though Eq. (\ref{z
term}) is written in terms of the slow roll parameters, it is an exact
result.     In    order    to    calculate    the    power    spectrum
$P_{\mathcal{R}}(k)$,  we  need  to solve Eq.
(\ref{mode eq 1}) and evaluate the quantity $|u_{k}/z|$ for every mode
with comoving  wavenumber $k$.   During inflation, these  modes evolve
from  a  quasi-Minkowskian state  that  can  be  represented at  short
wavelengths by the Bunch-Davies vacuum
\begin{equation}
\label{B-D vaccum 1}
u_{k} \propto e^{-ik \tau} \hspace{1cm} k\tau \rightarrow -\infty.
\end{equation}
When the  mode evolves to a  wavelength much greater  than the horizon
size, the solution is
\begin{equation}
\label{B-D vaccum 2}
u_{k} \propto z \hspace{1.5cm} k\tau \rightarrow 0.
\end{equation}
Equation (\ref{B-D vaccum 1}) together with the canonical quantization
condition for the fluctuations
\begin{equation}
\label{quantization condition}
u_{k}^{*} \frac{du_{k}}{d \tau}-u_{k}\frac{du_{k}^{*}}{d \tau}
=-i,
\end{equation}
completely specifies the initial conditions for the modes $u_{k}$.

Power-law  inflation, for  which $\epsilon=\eta=\xi=$  constant,  is a
case where Eq.  (\ref{mode eq 1}) can be solved
exactly,  with a  solution proportional  to a  Hankel function  of the
first kind:
\begin{equation}
\label{sol mode eq power law}
u_{k}\propto \sqrt{-k \tau}H_{\nu}(-k \tau),
\end{equation}
where
\begin{equation}
\label{nu power law}
\nu=\frac{3}{2}+\frac{\epsilon}{1-\epsilon}.
\end{equation}
For  the  slow  roll   approximation,  assuming  that  the  slow  roll
parameters $\epsilon$ and $\eta$ are small and approximately constant,
the solution  of the mode  equation is again given  by
Eq.  (\ref{sol mode eq power law}) with
\begin{equation}
\label{nu slow roll}
\nu=\frac{3}{2}+2\epsilon-\eta.
\end{equation}
Finally, the case of de Sitter expansion can be obtained from both the
power  law  and the  slow  roll results,  in  the  limit of  $\epsilon
\rightarrow 0$ and $\eta \rightarrow 0$.

We  can equivalently express  Eq. (\ref{mode  eq 1})  in terms  of the
variable $y$, which is defined as
\begin{equation}
\label{y}
y=\frac{k}{a H},
\end{equation}
and is the ratio of the wavelength of the mode relative to the horizon
size.  The variable $y$ is related to the conformal time $\tau$ by
\begin{equation}
\label{dy}
dy=-k(1-\epsilon)d\tau,
\end{equation}
and the mode equation (\ref{mode eq 1}) becomes
\begin{equation}
\label{mode eq y}
y^2(1-\epsilon)^2\frac{d^2u_{k}}{dy^2}+2y\epsilon(\epsilon-\eta)\frac{du_{k}}{dy}+\left[y^2-F(\epsilon, \eta, \xi)\right]u_{k}=0,
\end{equation}
where
\begin{equation}
\label{F}
F(\epsilon, \eta, \xi)=2\left(1+\epsilon+\epsilon^2-\frac{3}{2}\eta+\frac{1}{2}\eta^2-2\epsilon \eta+ \frac{1}{2}\xi^2 \right).
\end{equation}
A case  of special interest  in this paper  is the one where  the slow
roll  parameter  $\epsilon$  is  small  and  $\eta$  is  approximately
constant. The mode equation reduces then to
\begin{equation}
\label{mode eq y for small epsilon}
y^2\frac{d^2u_{k}}{dy^2}+ \left[y^2-(2-3\eta+\eta^2)\right]u_{k}=0,
\end{equation}
with the following normalized solution, 
\begin{equation}
\label{sol mode eq for small epsilon 1}
u_{k}= \frac{1}{2}\sqrt{\frac{\pi}{k}}[\sqrt{y}H_{\nu}(y)],
\end{equation}
where, for $\eta < 3/2$,
\begin{equation}
\nu = 3/2 - \eta.
\end{equation}

The  power spectrum  of curvature  perturbations is  evaluated  in the
long-wavelength limit,
\begin{equation}
\label{P(k) u small epsilon 1}
P^{1/2}_{\mathcal{R}}(k)=\sqrt{\frac{k^3}{2
\pi^2}}\left|\frac{u_{k}}{z}\right|_{y \rightarrow 0},
\end{equation}
which is given in terms of the variable $y$ by
\begin{equation}
\label{P(k) u small epsilon 2}
P^{1/2}_{\mathcal{R}}(k)=\mathcal{V}(\nu)\left(\frac{H^2}{2\pi\dot\phi} y^{3/2-\nu}\right),
\end{equation}
with
\begin{equation}
\label{mathcal V}
\mathcal{V}(\nu)=2^{\nu-3/2}\frac{\Gamma{(\nu)}}{\Gamma{(3/2)}}.
\end{equation}
The  scalar   spectral  index   $n$  is  defined   by  differentiating
Eq. (\ref{P(k) u small epsilon  2}) at constant time, or equivalently,
constant scale factor:
\begin{equation}
\label{spec ind for small epsilon 1}
n-1 \equiv \left. \frac{d \ln P_{\mathcal{R}}}{d \ln k}\right|_{aH=const.} = 3 - 2 \nu = 2 \eta.
\end{equation}

We emphasize  that the definition  of the curvature power  spectrum in
the  long-wavelength limit Eq.   (\ref{P(k) u  small epsilon  1}), can
differ substantially  from the widely used  definition of $P_{\mathcal
R}$ as evaluated at horizon crossing
\begin{equation}
\label{P(k) horizon crossing}
P^{1/2}_{\rm Hor}(k)=\sqrt{\frac{k^3}{2
\pi^2}}\left|\frac{u_{k}}{z}\right|_{y = 1}.
\end{equation}
In  the slow  roll  limit, Eqs.  (\ref{P(k)  u small  epsilon 1})  and
(\ref{P(k) horizon  crossing}) are equivalent  to second order  in the
slow  roll  parameters. However,  in  situations  where  slow roll  is
strongly  violated,  the horizon  crossing  expression  for the  power
spectrum cannot be used, and  care must be taken to correctly evaluate
the    power   spectrum   in    the   long-wavelength    limit.   (See
Ref. \cite{Kinney:2005vj}  for a detailed discussion  of the breakdown
of the horizon crossing formalism far from slow roll.)

In the  next section, we specialize  to a particular  case of non-slow
roll evolution, which corresponds to a field rolling over the top of a
local maximum of a potential.

\section{Inflation over the hill}

The specific example of non-slow  roll evolution we consider here is a
scalar  field  rolling   over  the  top  of  a   local  maximum  of  a
potential. This can  be seen to be intrinsically  non-slow roll, since
in the slow roll limit
\begin{equation}
\label{SR eq motion for small field}
\dot\phi \propto V'\left(\phi\right) = 0,
\end{equation}
at the  maximum of  the potential.  The  slow roll  solution therefore
automatically corresponds  to a solution with the  field stationary at
the maximum  of the  potential in the  infinite past.  In  a realistic
situation, for example  a field evolving in the  string landscape, the
field velocity may well be  nonzero near the maximum.  Since slow roll
inflation is  generically an attractor  solution \cite{Liddle:1994dx},
the evolution may still relax to the slow roll attractor at late times
even  if it  is far  from slow  roll at  early times.   

We consider a potential of the form
\begin{equation}
\label{small field V}
V(\phi)=\Lambda^4 \left[1-\frac{1}{2}\left(\frac{\phi}{\mu}\right)^2\right].
\end{equation}
If the mass term of the inflaton is unsuppressed, the above expression
will be a  good approximation for {\em any}  single-field potential as
long  as  the inflaton  is  sufficiently  close  to the  maximum.  The
equation of motion (\ref{EoM}) can then be written as
\begin{equation}
\label{small field EoM 1}
\ddot \phi + 3 H \dot \phi-\frac{\Lambda^4}{\mu^2}\phi =0.
\end{equation}
Assuming that $\dot  \phi \ll V(\phi)$ and $\phi  \ll \mu$, the Hubble
parameter $H$ can be treated as approximately constant:
\begin{equation}
\label{small field H}
H \approx \sqrt{\frac{8 \pi}{3m_{\rm Pl}^2}\Lambda^4}.
\end{equation}

Following  the  standard slow  roll  approximation,  the second  order
differential equation  of motion (\ref{small field EoM  1}) reduces to
the  first  order  Eq. (\ref{SR  EoM}),  with
$\dot\phi \propto V'\left(\phi\right)$.  But if one considers the full
equation  of  motion  (\ref{small  field  EoM 1}),  it  can  be  shown
\cite{Inoue:2001zt}  that the  evolution of  $\phi$ can  be decomposed
into a  slowly varying  branch, which is  identified as the  slow roll
solution, plus a rapidly  changing branch proportional to $a^{-3}(t)$.
In this paper we relax the  slow roll assumption and we study the full
equation of motion of the field.

Equation (\ref{small  field EoM 1}) can  be expressed in  terms of the
number of e-folds $N$ by using Eq.  (\ref{N}) as
\begin{equation}
\label{small field EoM 2}
\frac{d^2\phi}{dN^2} - 3 \frac{d\phi}{dN} -\alpha \phi=0, 
\end{equation}
where $\alpha$ is given by
\begin{equation}
\label{small field alpha}
\alpha=\frac{\Lambda^4}{\mu^2H^2} \approx \frac{3}{8\pi}\left(\frac{m_{\rm Pl}}{\mu}\right)^2.
\end{equation}
The  general solution  of  the equation  of  motion (\ref{small  field
EoM 2}) is then
\begin{equation}
\label{small field phi}
\phi(N)=\phi_{+} e^{r_{+}N}+\phi_{-} e^{r_{-}N},
\end{equation}
with
\begin{equation}
\label{small field full r+-}
r_{\pm}=\frac{3}{2}\left(1 \mp \sqrt{1+\frac{4}{9}\alpha} \right).
\end{equation}
If  $\alpha$  is  small, the  square  root  can  be expanded  and  the
constants $r_{\pm}$ become
\begin{eqnarray}
\label{small field approx r+-}
r_{+} & = & -\frac{\alpha}{3}, \nonumber\\
r_{-} & = & 3 + \frac{\alpha}{3}.
\end{eqnarray}

The general  solution (\ref{small field phi})  has two distinguishable
branches: a  transient branch $r_{-}$  which dominates at  early time,
and a ``slow  rolling'' branch $r_{+}$ which can  be identified as the
late-time attractor.   Since $\alpha$ is  positive definite, $r_{\pm}$
will always be  real, and an inflationary solution  will exist for any
set of chosen parameters $\Lambda^4$ and $\mu$.

In what follows,  we assume that the solution  (\ref{small field phi})
describes the evolution of the  field for some range of inflation, and
we  look  for  new  features  when  calculating  the  curvature  power
spectrum.  From  Eqs.  (\ref{small  field phi}) and  (\ref{small field
full r+-}) it is clear that the ratio of the transient branch over the
``slow rolling''  one decreases  exponentially.  That means  that both
branches will be equal at some value of the number of e-folds $N_{0}$.
We can  arbitrarily set $N_{0}$  by choosing the relative  strength of
the  parameters $\phi_{+}$  and  $\phi_{-}$ in  Eq. (\ref{small  field
phi}).   Despite the  fact that  the $r_{-}$  branch  is exponentially
decaying relative  to the $r_{+}$ branch, the  transient ``fast roll''
evolution can still correspond to an observationally relevant range of
scales. This can easily be  seen by considering the relation between the comoving wavenumber and the field value \cite{Liddle:1994cr}:
\begin{equation}
d(\ln{k})=\frac{2 \sqrt{\pi}}{m_{pl}}\frac{(\epsilon - 1)}{\sqrt{\epsilon}} d \phi.
\end{equation} 
Even
though  the  field displacement  which  corresponds  to the  transient
branch is small, in the limit that $\epsilon$ is small, the field
evolution  during  the transient can be mapped to a large range of comoving wavenumbers. This is also evident in the particular case shown in Fig. \ref{fig:eta}, where the transition between the transient and the attractor solution lasts about four e-folds of expansion.

In  the following two  subsections, we  calculate the  curvature power
spectrum for two cases.  We first consider each branch separately, for
which $\eta$  is approximately constant.  We then  calculate the power
spectrum in the region where the field evolves from the $r_{-}$ to the
$r_{+}$ branch and $\eta$ is evolving rapidly.

\subsection{{\boldmath $\phi=\phi_{\pm}e^{r_{\pm}N}$}}

For simplicity, we first consider the two solutions in
Eq. (\ref{small     field    phi})    separately.     Using
Eqs.  (\ref{intro eps  3})  and  (\ref{intro eta  5}),  the slow  roll
parameters are given by
\begin{eqnarray}
\label{A SR params}
\epsilon & = & 4 \pi r_{\pm}^2\left(\frac{\phi}{m_{\rm Pl}}\right)^2,\nonumber\\
\eta & = & r_{\pm}+\epsilon.
\end{eqnarray}
Near the maximum of the  potential, $\phi \approx 0$ and $\epsilon \ll
1$, so that 
\begin{equation}
\label{A eta 1}
\eta \approx r_{\pm}=constant.
\end{equation}
Even though $\epsilon$ is assumed to be small, it is obvious from Eqs.
(\ref{small field  approx r+-}) and  (\ref{A eta 1}) that  the $r_{-}$
branch  will never satisfy  the slow  roll conditions,  since $\eta>3$
always.  For the  $r_{+}$ branch, the slow roll  limit is obtained for
$\alpha \ll 1$ for which
\begin{equation}
\label{A eta 2}
|\eta_{SR}| \approx \frac{\alpha}{3} \ll 1.
\end{equation}
This is the  limit which will be of  observational interest, since the
late-time  attractor  corresponds to  a  nearly scale-invariant  power
spectrum consistent with observations.

The mode  equation (\ref{mode eq  y}) for the  curvature perturbations
can then be written in either limit as
\begin{equation}
\label{A mode eq y}
y^2\frac{d^2u_{k}}{dy^2}+[y^2-(2-3\eta+\eta^2)]u_{k}=0,
\end{equation}
where all the $\epsilon$ terms have been neglected. The above equation
is invariant  under the transformation $\eta  \rightarrow 3-\eta$, and
hence any pair $\eta_{1}$ and $\eta_{2}$ that satisfies
\begin{equation}
\label{A eta 1 2}
\eta_{2}=3-\eta_{1},
\end{equation}
will give the  same mode equation and therefore  the same solution. It
should be noted that the above invariance breaks down in the case of a
non-negligible  $\epsilon$.   Since  $\epsilon$ is  negligible,  $\dot
\phi$  is small and  therefore calling  the transient  solution ``fast
roll''  is  something of  a  misnomer, despite  the  fact  that it  is
not slow roll.

The  parameters  $r_{\pm}$ defined  in  Eq.  (\ref{small field  approx
r+-}), satisfy Eq.   (\ref{A eta 1 2}) and therefore  lead to the same
mode equation,  which can be  solved analytically.  The  solutions are
proportional to a Hankel function of the first kind
\begin{equation}
\label{A sol mode eq}
u_{k}\propto \sqrt{y} H_{\nu}(y),
\end{equation}
where
\begin{equation}
\label{A nu}
\nu=\left|\frac{3}{2}-\eta\right|.
\end{equation}
The absolute value of $\nu$ is a consequence of the fact that the mode
equation is  the same for both  branches.  This means  that the Hankel
functions  should  be  of the  same  order  in  both cases,  and  this
requirement is  satisfied by Eq.   (\ref{A nu}).  The  significance of
the  absolute  value can  also  be  seen  by considering  the  general
solution of the mode equation (\ref{A mode eq y}),
\begin{equation}
\label{general solution}
u_{k}=\frac{1}{2}\sqrt{\frac{\pi}{k}} \sqrt{y} [a_{k}H_{\nu}(y)+b_{k}H_{\nu}^{*}(y)]. 
\end{equation}
When the mode is well inside the horizon, the above equation becomes
\begin{equation}
\label{approximate solution}
u_{k} \propto a_{k}e^{iy}+b_{k}e^{-iy}.
\end{equation}
The  Bunch-Davies   vacuum  is  obtained  by   setting  $b_{k}=0$  and
$a_{k}=1$.  If  instead we  look  at  the  general solution  for  $\nu
\rightarrow -\nu$, we can use the identity
\begin{equation}
H_{-\nu} = e^{i \nu \pi} H_{\nu},
\end{equation}
to write
\begin{eqnarray}
&&u_{k}=\frac{1}{2}\sqrt{\frac{\pi}{k}} \sqrt{y} [ \tilde a_{k}H_{- \nu}(y)+ \tilde b_{k}H_{- \nu}^{*}(y)]\cr
&&=\frac{1}{2}\sqrt{\frac{\pi}{k}} \sqrt{y} [e^{i\nu \pi} \tilde a_{k}H_{\nu}(y) + e^{-i \nu \pi} \tilde b_{k}H_{\nu}^{*}(y)],
\end{eqnarray}
which corresponds  to the Bunch-Davies boundary  condition for $\tilde
a_k  =  e^{-i  \nu  \pi}$  and  $\tilde  b_k  =  0$.   Therefore  $\nu
\rightarrow -\nu$ introduces an  irrelevant overall phase shift.  This
corresponds  precisely to the  symmetry of  the mode  equation (\ref{A
mode eq y}) under the transformation $\eta \rightarrow 3 - \eta$.

The power  spectrum of curvature perturbations can  be calculated from
Eq. (\ref{P(k) u small epsilon 2})
\begin{equation}
\label{P(k) u small epsilon 3}
P_{\mathcal R}^{1/2} \propto k^{3/2} \left\vert u_k \over z\right\vert_{y \rightarrow 0} \propto \frac{H^2}{2\pi \dot\phi} y^{3/2-\nu},
\end{equation}
and since  $\nu$ takes  the same value  for both the  $\eta=r_{+}$ and
$\eta=r_{-}$ cases,
\begin{equation}
\label{A pow spec 2}
P_{\mathcal R}^{1/2}  \propto \frac{H^2}{2 \pi \dot \phi}y^{r_{+}},
\end{equation}
for both branches. The scalar spectral index is then given by
\begin{equation}
\label{A spec ind 1}
n-1=\left. \frac{d \ln P_{\mathcal R}}{d\ln k}\right|_{aH={\rm const.}}=2r_{+},
\end{equation}
which is the  first order slow roll expression  for the spectral index
in  the  small  $\epsilon$  limit  for  the  slowly  evolving  $r_{+}$
branch. From Eq.  (\ref{small field  approx r+-}) it is clear that the
choice $\alpha=0.075$ corresponds to
\begin{equation}
n=0.95,
\end{equation}
which   is   within   the   region   favored   by   the   WMAP3   data
\cite{Spergel:2006hy,Alabidi:2006qa,Seljak:2006bg,Kinney:2006qm,Martin:2006rs}.

Even though therefore the field may  evolve in a region where the slow
roll approximation is strongly  violated ($\eta \rightarrow 3^+$), the
resultant power spectrum is identical to the one given by an evolution
that satisfies the slow roll approximation ($\eta \rightarrow 0^-$) as
a  result   of  the  invariance   of  the  mode  equation   under  the
transformation  given  by Eq.   (\ref{A  eta  1  2}).  Note  that  the
$\eta=3$   case    corresponds   to   ultra    slow   roll   inflation
\cite{Tsamis:2003px},   which   gives   an   exact   scale   invariant
spectrum. Evaluating the mode function at horizon crossing, results in
an  incorrect   expression  for  the  shape  of   the  power  spectrum
Eq. (\ref{P(k) horizon crossing}), since
\begin{equation}
n_{\rm Hor} - 1 = \left. \frac{d \ln P_{\mathcal R}}{d\ln k}\right|_{k = a H}=2 \eta = 2 r_{-}.
\end{equation}

Finally, it can be shown that the transient $r_{-}$ branch describes a
field rolling  up an inverted potential.  Re-expressing the derivative
of the  potential with  respect to the  field in Eq.   (\ref{intro eta
2}), as a derivative with respect to the number of e-folds
\begin{equation}
\label{eta for small epsilon}
\eta=3-\frac{1}{\dot \phi^2}\frac{dV(\phi)}{dN}.
\end{equation}
For the $r_{-}$ branch, $\eta > 3$, so that
\begin{equation}
\label{dV / dN}
\frac{dV(\phi)}{dN}<0,
\end{equation}
and since $dN<0$  during inflation, we find that  the field is rolling
up the potential,
\begin{equation}
\label{dV }
\Delta V(\phi)>0.
\end{equation}

We next  consider the  transition from the  transient solution  to the
late-time attractor.

\subsection{{\boldmath $\phi=\phi_{+}e^{r_{+}N} +\phi_{-}e^{r_{-}N}$}}

In the previous  section we showed that the  transient $r_{-}$ branch,
even though  it describes inflationary  solutions far form  slow roll,
generates   curvature   perturbations   that   are   consistent   with
observations as a result of  the invariance of the mode equation under
the  transformation  (\ref{A  eta  1  2}).   These  perturbations  are
identical  to the ones  produced by  the $r_{+}$  branch, and  in that
sense the two branches  are indistinguishable in terms of observables,
since they  both produce negligible gravitational  waves.  The $r_{-}$
branch,  however,  corresponds  to  the  intrinsically  non-slow  roll
solution of a field rolling up  the hill of an inverted potential.  In
this section,  we investigate the  power spectrum produced  during the
transition  from the transient  $r_{-}$ branch  to the  late-time slow
roll  solution  represented  by   the  $r_{+}$  branch.   During  this
transition, the parameter $\eta$  varies rapidly, and it is reasonable
to  expect  a  feature  in  the curvature  power  spectrum  at  scales
corresponding  to that transition.  Remarkably, we  find that  no such
feature exists.

In order  for the field  $\phi$ to be  monotonic in time,  $\dot \phi$
must not  change sign. This requirement  can be expressed  in terms of
$N$ as follows
\begin{equation}
\label{B dphi dN}
\frac{d\phi}{dN}=r_{+}\phi_{+}e^{r_{+}N}+r_{-}\phi_{-}e^{r_{-}N}\neq0,
\end{equation}
assuming that  the derivative of  the field with  respect to $N$  is a
smooth  function throughout  its evolution.   Since the  signs  of the
constants $r_{\pm}$ are known  from Eq.  (\ref{small field full r+-}),
the above  condition can be achieved  only if one  of the coefficients
$\phi_{\pm}$  is negative. We  take $\phi_{-}<0$,  which results  in a
positive time derivative for the field
\begin{equation}
\label{B dphi dt}
\dot \phi=-H\frac{d\phi}{dN}>0.
\end{equation}
Using Eqs. (\ref{intro eps 3})  and (\ref{intro eta 5}), the slow roll
parameters are then given exactly by
\begin{eqnarray}
\label{B SRparams}
\epsilon&=&\frac{4\pi}{m_{\rm Pl}^2}\left(r_{+}\phi_{+}e^{r_{+}N}+r_{-}\phi_{-}e^{r_{-}N}\right)^2,\nonumber\\
\eta&=&\frac{r_{+}^2\phi_{+}e^{r_{+}N}+r_{-}^2\phi_{-}e^{r_{-}N}}{r_{+}\phi_{+}e^{r_{+}N}+r_{-}\phi_{-}e^{r_{-}N}}+\epsilon,
\end{eqnarray}
where in the small $\epsilon$ limit
\begin{equation}
\label{B eta 1}
\eta \approx \frac{r_{+}^2\phi_{+}e^{r_{+}N}+r_{-}^2\phi_{-}e^{r_{-}N}}{r_{+}\phi_{+}e^{r_{+} N}+r_{-}\phi_{-} e^{r_{-}N}}.
\end{equation}
According to the above sign convention for $\phi_{-}$, the denominator
in Eq. (\ref{B eta 1}) never crosses zero and $\eta$ is always finite.
Equation (\ref{B eta 1}) also reproduces both Eqs.  (\ref{A eta 1}) in
the limiting case where one of the two branches vanishes.

Expressing the slow roll parameter  $\xi^2$ in terms of $\epsilon$ and
$\eta$ as \cite{Kinney:2002qn}
\begin{equation}
\label{xi^2 2}
\xi^2=\frac{d\eta}{dN}+\epsilon \eta,
\end{equation}
and  ignoring all  the terms  of order  $\epsilon$, the  mode equation
becomes
\begin{equation}
\label{B mode eq}
y^2\frac{d^2u_{k}}{dy^2}+\left[y^2-F(\eta)\right]u_{k}=0,
\end{equation}
where
\begin{equation}
\label{F(nu)}
F(\eta)=2-3\eta+\eta^2+\frac{d\eta}{dN}.
\end{equation}
Equation  (\ref{B   mode  eq})  is  no  longer   invariant  under  the
transformation (\ref{A eta 1 2}) since
\begin{equation}
\label{F(nu) not inv}
F(\eta)\neq F(3-\eta).
\end{equation}
Even though the mode equation  does not respect the above symmetry, it
can be solved analytically.  In order  to see that, we first note that
both branches will be equal at some number of e-folds $N_{0}$
\begin{equation}
\label{equalbranches}
\phi_{+}e^{r_{+}N_{0}}=-\phi_{-}e^{r_{-}N_{0}}.
\end{equation}
Equation (\ref{small field phi}) then becomes
\begin{equation}
\label{B phi}
\phi=A\left[e^{r_{+}(N-N_{0})}-e^{r_{-}(N-N_{0})}\right],
\end{equation}
where
\begin{equation}
\label{small field A}
A=\phi_{+}e^{r_{+}N_{0}}=-\phi_{-}e^{r_{-}N_{0}}.
\end{equation}
We are therefore able to absorb the coefficients $\phi_{\pm}$ into $A$
using the  equality of the branches at  $N=N_{0}$, and
Eq. (\ref{B eta 1}) takes then the simpler form
\begin{equation}
\label{B eta 2}
\eta=\frac{r_{+}^2e^{r_{+}(N-N_{0})}-r_{-}^2e^{r_{-}(N-N_{0})}}{r_{+}e^{r_{+}(N-N_{0})}-r_{-}e^{r_{-}(N-N_{0})}}.
\end{equation}
Figure 1 shows $\eta$ as a  function of $N$ for the case of $N_{0}=58$
and $\alpha=0.075$, where it can  be seen that the transition from the
$r_{-}$ to the $r_{+}$ branch lasts for approximately four e-folds.

\begin{figure}[!h]
\includegraphics[angle=270, width=\columnwidth] {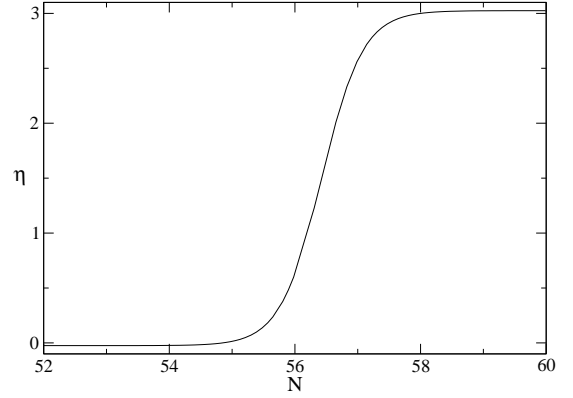}
\caption{$\eta$   as   a   function   of  $N$   for   $N_{0}=58$   and
$\alpha=0.075$}
\label{fig:eta}
\end{figure}
Using Eq. (\ref{B eta  2}) it can also be shown
that
\begin{equation}
\label{B deta dN}
\frac{d\eta}{dN}=-\frac{r_{+}r_{-}(r_{+}-r_{-})^2e^{(N-N_{0})(r_{+}+r_{-})}}{\left[r_{+}e^{r_{+}(N-N_{0})}-r_{-}e^{r_{-}(N-N_{0})}\right]^2}.
\end{equation}
We  therefore have analytic  expressions for  the slow  roll parameter
$\eta$ and its derivative.  Substituting back these two equations into
Eq. (\ref{F(nu)}), we  can find  an exact  expression  for the
function $F(\eta)$.   After some straightforward  but tedious algebra,
and using  the full expressions  (\ref{small field full r+-})  for the
parameters $r_{\pm}$ it can be shown that
\begin{equation}
\label{B F(nu) 2}
F(\eta)=2+\alpha,
\end{equation}
for any value  of $N_{0}$.  Even though $\eta$  evolves rapidly during
the  transition between  the two  branches (Fig.   \ref{fig:eta}), the
function  $F(\eta)$ remains  constant, and  Eq.
(\ref{B mode eq}) becomes
\begin{equation}
\label{B mode eq final}
y^2\frac{d^2u_{k}}{dy^2}+\left[y^2-(2+\alpha)\right]u_{k}=0.
\end{equation}
This invariance of the $z''/z$ term in the mode equation is an example of the duality invariance considered by Wands in Ref. \cite{Wands:1998yp}, and a non-slow roll solution exhibiting the same invariance was obtained by Leach and Liddle in Ref. \cite{Leach:2000yw} for a different choice of potential. More recently, Starobinsky \cite{Starobinsky:2005ab} has constructed a non-slow roll solution resulting in an exactly scale-invariant spectrum. 

The solution of Eq. (\ref{B mode eq final}) can be written in terms of a Hankel function
\begin{equation}
\label{B sol mode eq final}
u_{k} \propto \sqrt{y} H_{\nu}(y),
\end{equation}
where
\begin{equation}
\label{B nu}
\nu=\frac{3}{2}\sqrt{1+\frac{4}{9}\alpha},
\end{equation}
and the power spectrum of curvature perturbations
\begin{equation}
\label{B pow spec 1}
P_{\mathcal{R}}^{1/2} \propto \left|\frac{u_{k}}{z}\right|_{y \rightarrow 0} \propto \frac{H^2}{2 \pi \dot \phi}y^{3/2-\nu},
\end{equation}
using  Eqs. (\ref{small field  full r+-})  and (\ref{B  nu}) becomes
\begin{equation}
\label{B pow spec 2}
P_{\mathcal{R}}^{1/2} \propto \frac{H^2}{2\pi \dot \phi}y^{r_{+}}.
\end{equation}
The scalar spectral index is then given by
\begin{equation}
\label{B spec ind}
n-1=\left. \frac{d \ln P_{k}}{d\ln k}\right|_{aH=const.}=2r_{+}.
\end{equation}
We therefore  obtain a slightly  red spectrum since $r_{+}<0$  with no
running
\begin{equation}
\label{running}
\frac{dn}{d \ln k}=0.
\end{equation}
The results (\ref{B pow spec  2}) and (\ref{B spec ind}) are identical
to the  results (\ref{A pow spec 2})  and (\ref{A spec ind  1}) of the
previous  section.   Even though  the  transition  from the  transient
$r_{-}$  branch to  the  slowly evolving  late-time attractor  $r_{+}$
branch results  in a rapidly  varying slow roll parameter  $\eta$, the
mode  equation  can be  solved  exactly  and  the resultant  curvature
spectrum  is an  exact power  law spectrum. Despite the fact that allowing for the transient solution adds a parameter $N_0$ to the model, the observable power spectrum is independent of that additional parameter. Thus the usual correspondence in slow roll between the field value and the wavenumber of the fluctuation is broken, but without an effect on the power spectrum.

The above  analysis assumes that $\epsilon$  is negligible throughtout
the relevant range of scales  and therefore can be neglected. In order
to  check   this  assumption,  the  full  mode   equation  was  solved
numerically  for  an  appropriate  set  of  initial  conditions  which
correspond  to $\epsilon  \approx 10^{-4}$  at N=65.   This  value was
chosen so that modes that  become superhorizon at N=60 are well within
the  horizon at  N=65. Since  the field  is rolling  up  the potential
initially, its  velocity is  suppressed dramatically making  the above
assumption an  excellent approximation, giving a  value for $\epsilon$
which is smaller by more that  10 orders of magnitude at N=60. For these values of $\epsilon$, the analytic and numerical results match to ${\delta n} / {n} < 10^{-5}$, and we can  therefore safely neglect  all the $\epsilon$ terms  and obtain the exact solution (\ref{B sol mode eq final}).

We can  also calculate  the potential which  corresponds to  the above
field evolution.  From Eqs.  (\ref{HJ}) and (\ref{intro eps 1}) it can
be shown that,
\begin{equation}
\label{B pot}
V(\phi)=\frac{3m_{\rm Pl}^2}{8
\pi}H^2(\phi)\left[1-\frac{1}{3}\epsilon(\phi)\right].
\end{equation}
Taking the derivative with respect to $\phi$ and using Eqs.  (\ref{N})
and (\ref{intro eps 4})
\begin{equation}
\label{B dpot dN 1}
\frac{dV}{dN}=\frac{3m_{\rm Pl}^2}{8\pi}H^2\left(2\epsilon-\frac{2}{3}\epsilon^2 -\frac{1}{3} \frac{d\epsilon}{dN}\right),
\end{equation}
where \cite{Kinney:2002qn}
\begin{equation}
\label{B deps dN}
\frac{d\epsilon}{dN}=2 \epsilon(\eta-\epsilon).
\end{equation}
Substituting the above expression into Eq. (\ref{B dpot dN 1}),
\begin{equation}
\label{B dpot dN 2}
\frac{dV}{dN}=\frac{3m_{\rm Pl}^2}{4\pi}\epsilon H^2\left(1-\frac{\eta}{3}\right).
\end{equation}
Equation (\ref{B dpot dN 2}) gives the same qualitative results as Eq.
(\ref{eta for  small epsilon}).  Early on, when  the transient $r_{-}$
branch dominates,
\begin{equation}
\label{case 1}
\eta>3 \Longrightarrow \Delta V(\phi)>0,
\end{equation}
and the field is rolling  up the potential. Later during its evolution
along $V$,
\begin{equation}
\label{case 2}
\eta<3 \Longrightarrow \Delta V(\phi)<0,
\end{equation}
and the field is rolling  down the potential.  We have therefore shown
that  a field  can  evolve on  both  sides around  the  maximum of  an
inverted potential  and generate  curvature perturbations that  are in
perfect agreement  with observations.  

We can trivially  extend the above analysis to  the case of tree-level
hybrid      inflation      \cite{Linde:1993cn}      considered      in
Refs.  \cite{Garcia-Bellido:1996ke,Kinney:1997ne,Kinney:2005vj}  where
the potential is given by
\begin{equation}
\label{hybrid V}
V(\phi)=M^4+\frac{1}{2}\mu^2\phi^2.
\end{equation}
Following the procedure  of the previous section it  can be shown that
the equation of motion for the field in terms of the number of e-folds
$N$ can be written as
\begin{equation}
\label{hybrid EoM}
\frac{d^2\phi}{dN^2} - 3 \frac{d\phi}{dN} +\alpha \phi=0, 
\end{equation}
where
\begin{equation}
\label{hybrid alpha}
\alpha=\left(\frac{\mu}{H}\right)^2 \approx \frac{3}{8\pi}\left(\frac{m_{\rm Pl}\mu}{M^2}\right)^2,
\end{equation}
and 
\begin{equation}
\label{hybrid H}
H \approx \sqrt{\frac{8 \pi}{3m_{\rm Pl}^2}M^4}= {\rm const.}
\end{equation}
The general solution  of the equation of motion  (\ref{hybrid EoM}) is
then
\begin{equation}
\label{hybrid phi}
\phi(N)=\phi_{+}e^{r_{+}N}+\phi_{-}e^{r_{-}N},
\end{equation}
where 
\begin{equation}
\label{hybrid full r+-}
r_{\pm}=\frac{3}{2}\left(1 \mp \sqrt{1-\frac{4}{9}\alpha} \right).
\end{equation}
Comparing the  above equation with Eq.  (\ref{small  field full r+-}),
it can be seen that there  is only a sign difference between them.  In
order  therefore  for  the   above  solution  to  describe  inflation,
$r_{\pm}$  must be  real, which  can  be translated  in the  following
condition for the parameters of the corresponding potential
\begin{equation}
\label{condition for aplha}
\alpha<\frac{9}{4}.
\end{equation}
In the small $\alpha$ limit
\begin{eqnarray}
\label{hybrid approx r+-}
r_{+} & = & \frac{\alpha}{3}, \nonumber\\
r_{-} & = & 3 - \frac{\alpha}{3}.
\end{eqnarray}
Since $r_{\pm}>0$, the requirement  that $\phi$ should be monotonic in
time  is satisfied only  if both  coefficients $\phi_{\pm}$  have the
same sign.

In Ref.  \cite{Kinney:2005vj}, WHK  showed that both branches give the
same  power  spectrum  of curvature  perturbations  $P_{\mathcal{R}}$.
Following  the  argument  presented   in  the  previous  section,  the
constants $r_{\pm}$  defined in Eq. (\ref{hybrid  approx r+-}) satisfy
Eq.   (\ref{A eta  1 2})  and therefore  correspond to  the  same mode
equation.   The power  spectrum and  the spectral  index will  then be
given  by  Eqs.   (\ref{A  pow  spec  2}) and  (\ref{A  spec  ind  1})
respectively, where $r_{+}$  is positive in this case,  resulting in a
blue spectrum.

We can study  the more general case of  the full equation (\ref{hybrid
phi})  for the  field evolution  using the  analysis for  the inverted
potential, by making the substitutions
\begin{eqnarray}
\label{substitutions}
-\phi_{-} & \rightarrow & \phi_{-} \nonumber\\
\alpha &  \rightarrow & -\alpha. 
\end{eqnarray}
Equation (\ref{hybrid phi}) can then be written
\begin{equation}
\label{hybrid phi 2}
\phi=A\left[e^{r_{+}(N-N_{0})}+e^{r_{-}(N-N_{0})}\right],
\end{equation}
where
\begin{equation}
\label{hybrid A}
A=\phi_{+}e^{r_{+}N_{0}}=\phi_{-}e^{r_{-}N_{0}}.
\end{equation}
Equations (\ref{B eta 2}) and (\ref{B deta dN})
take also the form
\begin{equation}
\label{hybrid eta}
\eta=\frac{r_{+}^2e^{r_{+}(N-N_{0})}+r_{-}^2e^{r_{-}(N-N_{0})}}{r_{+}e^{r_{+}(N-N_{0})}+r_{-}e^{r_{-}(N-N_{0})}},
\end{equation}
and
\begin{equation}
\label{hybrid deta dN}
\frac{d\eta}{dN}=\frac{r_{+}r_{-}(r_{+}-r_{-})^2e^{(N-N_{0})(r_{+}+r_{-})}}{\left[r_{+}e^{r_{+}(N-N_{0})}+r_{-}e^{r_{-}(N-N_{0})}\right]^2},
\end{equation}
respectively. Substituting back these expressions into Eq.
(\ref{F(nu)}), we find that
\begin{equation}
\label{hybrid F(nu)}
F(\eta)=2-\alpha,
\end{equation}
and the mode equation becomes
\begin{equation}
\label{hybrid mode eq}
y^2\frac{d^2u_{k}}{dy^2}+\left[y^2-(2-\alpha)\right]u_{k}=0.
\end{equation}
The solution is again a Hankel function
\begin{equation}
\label{hybrid sol mode eq}
u_{k} \propto \sqrt{y} H_{\nu}(y),
\end{equation}
with
\begin{equation}
\label{hybrid nu}
\nu=\frac{3}{2}\sqrt{1-\frac{4}{9}\alpha}.
\end{equation}
The  power spectrum of  curvature perturbations  $P_{\mathcal{R}}$ and
the spectral  index $n$ are then  given by the Eqs.   (\ref{B pow spec
2}) and (\ref{B  spec ind}) respectively, where $r_{+}$  is defined by
Eq.  (\ref{hybrid  full r+-}) or  by (\ref{hybrid approx r+-})  in the
small $\alpha$ limit.

\section{Conclusions}

In this paper we considered the  equation of motion for the case of an
inverted potential given by  Eq. (\ref{small field V}) with the
general solution
\begin{equation}
\label{conclusion small field phi}
\phi(N)=\phi_{+}e^{r_{+}N}+\phi_{-}e^{r_{-}N},
\end{equation}
which  consists of  two different  branches, the  early-time transient
$r_{-}$  branch and  the  slowly rolling  late-time attractor  $r_{+}$
branch.   For simplicity,  we  initially considered  the two  branches
separately.  The $r_{-}$ branch describes inflationary solutions which
are always  far from slow roll,  corresponding to values  for the slow
roll parameter $\eta>3$.  Specifically, in the small $\epsilon$ limit,
\begin{eqnarray}
\label{conclusion small field approx r+-}
\eta & = & 3 + \frac{\alpha}{3}, \nonumber\\
\eta & = &-\frac{\alpha}{3},
\end{eqnarray}
for  the  $r_{-}$ and  the  $r_{+}$  branches
respectively,  where $\alpha$ is  a positive-definite  parameter which
depends on  the chosen potential.  The  shape of the  potential can be
derived from the behavior of  the slow roll parameter $\eta$ using the
following exact result
\begin{equation}
\label{conclusion eta for small epsilon}
\eta=3-\frac{1}{\dot \phi^2}\hspace{0.1cm}\frac{dV(\phi)}{dN}.
\end{equation}
Comparing  Eqs.    (\ref{conclusion  small  field   approx  r+-})  and
(\ref{conclusion eta for small epsilon}) we concluded that the $r_{-}$
branch describes a field rolling up an inverted potential since
\begin{equation}
\label{conclusion dV}
\Delta V(\phi)>0,
\end{equation}
while  the slow  roll  $r_{+}$  branch corresponds  to  the late  time
attractor evolution for the field.

Neglecting terms of order  $\epsilon$, the mode equation for curvature
perturbations becomes
\begin{equation}
\label{conclusion mode eq y}
y^2\frac{d^2u_{k}}{dy^2}+[y^2-(2-3\eta+\eta^2)]u_{k}=0,
\end{equation}
which is invariant under the transformation
\begin{equation}
\label{conclusion eta 1 2}
\eta \rightarrow 3-\eta.
\end{equation}
The  above transformation  is  exactly the  relation  between the  two
branches as can be seen  from Eqs. (\ref{conclusion small field approx
r+-}) which means  that both branches give the  same mode equation and
therefore  generate identical  curvature perturbations.   The spectral
index is given on both the $\eta=r_{+}$ and $\eta=r_{-}$ branches by
\begin{equation}
\label{conclusion spec ind}
n -1= 2r_{+},
\end{equation}
which corresponds in the slow roll  limit to a slightly ``red'', or $n
< 1$ power  spectrum, consistent with the region  favored by the WMAP3
data
\cite{Spergel:2006hy,Alabidi:2006qa,Seljak:2006bg,Kinney:2006qm,Martin:2006rs}.

We then considered the full equation for the field evolution
\begin{equation}
\label{conclusion phi}
\phi=\phi_{+}e^{r_{+}N}+\phi_{-}e^{r_{-}N}.
\end{equation}
In  this case  the slow  roll parameter  $\eta$ is  not constant  as a
result of the mixing of the two branches and the
mode equation becomes
\begin{equation}
\label{cosnclusion mode eq y 2}
y^2\frac{d^2u_{k}}{dy^2}+\left[y^2-\left(2-3\eta+\eta^2+\frac{d\eta}{dN}
\right)\right]u_{k}=0,
\end{equation}
where
\begin{equation}
\label{conclusion eta}
\eta=\frac{r_{+}^2e^{r_{+}(N-N_{0})}-r_{-}^2e^{r_{-}(N-N_{0})}}{r_{+}e^{r_{+}(N-N_{0})}-r_{-}e^{r_{-}(N-N_{0})}}
\end{equation}
and 
\begin{equation}
\label{conclusion deta dN}
\frac{d\eta}{dN}=-\frac{r_{+}r_{-}(r_{+}-r_{-})^2e^{(N-N_{0})(r_{+}+r_{-})}}{\left[r_{+}e^{r_{+}(N-N_{0})}-r_{-}e^{r_{-}(N-N_{0})}\right]^2}.
\end{equation}
Remarkably, the  quantity inside the parentheses in  the mode equation
is constant
\begin{equation}
\label{cosnclusion eta function}
2-3\eta+\eta^2+\frac{d\eta}{dN}=2+\alpha,
\end{equation}
and hence it can be solved exactly as
follows
\begin{equation}
\label{conclusion sol mode eq final}
u_{k} =\frac{1}{2}\sqrt{\frac{\pi}{k}} [\sqrt{y} H_{\nu}(y)],
\end{equation}
where
\begin{equation}
\label{conclusion nu}
\nu=\frac{3}{2}\sqrt{1+\frac{4}{9}\alpha}.
\end{equation}
The curvature power spectrum is therefore  given by a power law with a
red spectrum in {\em all} regions,
including  the  transition from  the  $r_{-}$  branch  to the  $r_{+}$
branch, with no features in the power spectrum.

Summarizing,  we constructed  a model  which in  the  small $\epsilon$
limit corresponds  to the inflaton field  rolling over the  hill of an
inverted  potential. Even  though the  slow roll  parameter  $\eta$ is
evolving rapidly  during the transition from  the early-time transient
to the late-time  attractor, the mode equation can  be solved exactly,
giving  curvature perturbations  which are  in perfect  agreement with
observations.  This introduces a much richer dynamical phase space for
inflationary model building, and may be useful for constructing models
on  the  string  landscape,  where  the total  number  of  e-folds  of
inflation  tends  to  be   small,  and  dynamical  transients  may  be
important.

\begin{center}
{\bf ACKNOWLEDGMENTS}
\end{center}
KT thanks  Brian Powell for  many useful discussions  and acknowledges
the support  of the Frank  B. Silvestro scholarship. This  research is
supported  in part  by  the National  Science  Foundation under  grant
NSF-PHY-0456777.

\end{document}